# The way forward
(Malcolm Fridlund, Artie Hatzes, René Liseau)

Our Solar System has a complex architecture that contains 1) Planets of several types, from giant gas giants to rocky terrestrial planets, 2) Debris disks, e.g. the zodiacal dust belt and the Kuiper-Edgeworth belt in the Solar System or disks of the β Pic or AU Mic type, and 3) Planetesimals or remnants thereof – e.g. Comets, Asteroids or Dwarf planets. There is also the Oort cloud of comets stretching tens of thousands of astronomical units from the Sun, but the understanding of the structure of the Oort cloud in the context of star formation is a separate issue and will not be discussed here. One of the ultimate goals of the research into both exoplanets and circumstellar disks is to be able to place this complex picture of our own Solar System into a proper context.

Answering questions like "Is our system a run-of-the-mill kind of system" or is it of a "rarer" kind – or even "unique" is of course central to this research. The answer to such questions is related to the overarching issue of whether we are alone in the Universe. The question of whether life exists elsewhere, than on the Earth, has been discussed by mankind since the beginning of recorded history, or even before, and it is still a fundamental question among scientists and laypersons alike today[1].

The understanding of how disks and planetary bodies relate to each other is the key to how planetary systems form and evolve. What kind of planets is the more common? What is the architecture of these planetary systems? And under what conditions were they formed? It must be a consequence of the formation- and evolution-processes that lead to the structure and physics of a specific system or an individual planet. This is very likely due to the conditions leading up to the formation of an accreting star and its associated disk. We can be fairly certain that the existence of planetary systems is a consequence of the process that forms stars. This understanding has not changed significantly since the time when we only had knowledge about only one such system – our own.

Our own solar system is of course very important since it is the only place in the Universe where we know that life exists. It is also the only system where we can study individual objects in situ and in exquisite detail. This is evidenced by the fact that we have more than half a dozen working spacecrafts in orbit or on the surface of Mars. As of this writing the New Horizons mission had just made its flyby of the dwarf planet Pluto on its way out into the Solar Systems outer debris disk – the Kuiper-Edgeworth belt. The pioneering Voyager 2 spacecraft has already crossed the heliopause and has become the first human spacecraft to be traveling in interstellar space. And, for more than a year the Rosetta spacecraft has been flying in tandem with Comet 67P/Churyumov-Gerasimenko, giving us valuable, in situ knowledge of a comet/planetesimal. These are indeed exciting times for solar system research.

---

[1] (as demonstrated by e.g. The recent announcement that "the Breakthrough Prize Foundation", an organization created by investor Yuri Milner, has announced that it will commit $100 million over the next ten years to a new, large-scale SETI initiative)

The understanding of star forming and debris disks of different structures and sizes is thus the key to understanding planet formation, specifically the types and masses of planets and the range of orbital distances that can form. Not only must a planet be located in the Habitable Zone (HZ, usually defined as the range of distance from the host star within liquid water can exist). Once such planets have formed, their orbits must be stable for a long enough time to provide the physical conditions for the formation of life.

Eventually, the central issue we wish to address is whether we are indeed alone in the Universe. The uniqueness or commonality of the Earth is fundamental to a proper understanding of such central issues as the properties needed by a planet in order for life to arise, the origin and evolution of life, and of course how common life is in the Universe. To correctly understand these issues we must also understand the context within which life on Earth exists and our place in the Universe. This thought, which has been with us throughout recorded history, can now be explored within the structure of a scientific inquiry. Is the Earth the only one of uncountable planets where life has arisen and evolved to the point that one of its life forms has the ability to formulate such a question? Or are the Earth, and its humans unique? Possibly the Earth formed out of random processes that were quite rare and very special circumstances were needed for higher life forms to evolve. The conditions may have been extremely rare and unlikely, but given the shear number of stars in our Galaxy perhaps it can occur at least once every several billions of years. It could also be that we may be alone in our galaxy, but certainly not alone in the vast Universe. Or….

The answer to such questions should be based on sound scientific inquiry and not just speculation.

**<u>The study of disks.</u>**

Interstellar disks and the process of stellar and planetary formation and evolution have now been studied in detail for more than 60 years. Through advances in Infrared and mm/sub-mm astronomy a consistent picture has slowly been built up. The Milky Way Galaxy, like all galaxies, is made up of stars, stellar remnants, planets, and an interstellar medium. This InterStellar Medium (ISM) is composed of atomic and molecular gas as well as solid matter in the form of dust grains. The ISM is mainly located in the plane of our Galaxy and it has different components, one of which is (giant) molecular clouds with a much higher ($10^3 – 10^4$ times higher) density than the "general" diffuse medium. These molecular clouds consist mainly of $H_2$, but with a significant enrichment of metals that supposedly originated in stars during the later stages of stellar evolution. These were then ejected into interstellar space through mass loss and supernova explosions to form the parental environment benevolent to the forming of stars and planets. While elements of the clouds collapse into protostars, some of the molecular material forms a protoplanetary disk, due to the conservation of angular momentum. This momentum preserves an enormous amount of rotational energy that the protostar has to lose before it can settle into a stable configuration. This is done through bipolar outflows (atomic and molecular), jets, as well as by the formation of planetary bodies.

Although most of the disk mass and its angular momentum is lost from the system through these processes, a large amount of material is nevertheless left behind in remnant  disks. This remaining material disperses slowly during the stellar lifetime either through collisions that form dust and debris disks or through ejections via interactions with the planetary bodies. However some of these debris disks remain as evidenced by the dust and large number of small bodies in the Solar System. We have begun to observe more and more of this material in other stellar systems (Eiroa et al., 2013) and essentially all of the stages outlined above have been detected either from the ground or space. Most recently the ESA space observatory Herschel has observed many aspects of these processes for the first time and in greater detail than ever before. On the ground the "first light" of the Atacama Large MM/sub-mm Array (ALMA) have detected the gaps in protoplanetary disks presumably resulting from the "sweeping up" of material by planets being formed (Partnership ALMA, 2015). While many of the details in these processes remain to be observed and clarified it appears that at least the broad elements of the picture are understood.

One important remaining step is to connect the understanding of the distribution of exoplanets (type of planets, type of orbits, type of host stars) with this picture of the early stages of a solar system. As is believed today (see below) planets are extremely common, and if not ubiquitous, anyhow important somehow to the star formation process. Nevertheless, one of the surprising results of the recent large surveys of exoplanets (e.g. ground based and/or space) is their diversity concerning radii, masses, orbital characteristics and their distribution among types of stars. The extreme dynamism of exoplanetary systems is also not what one expects from our understanding of the Solar System. According to numerical simulation the Solar System has remained stable enough for life to develop on the Earth, for more than 4 Gyrs although the major planets may have moved significantly with even Uranus and Neptune exchanging their order (Gomes et al., 2005) and it may remain so for at least as long although there is a finite possibility that the orbits of either Mars or Mercury (or both) may go haywire in the distant future with potentially disastrous consequences for the Earth (Batygin & Laughlin, 2008; Laskar, 2008). This stability may or may not be the norm as what concerns other planetary systems. In order to understand these facts it will be necessary to bring the picture of how stars and planets form together with what is actually observed of exoplanetary systems – never forgetting that we need to explain our own Solar System in the same picture.

**The study of exoplanets.**

 The issue of how common planets are in our Galaxy and by inference in the Universe was totally open until about 20 years ago. While the first 'modern' proposal for a 'proper' search for exoplanets date to more than 60 years ago (Struve, 1952), the first discoveries of bodies with masses comparable to those within the Solar System were made during the late 1980s and beginning of the 1990s,  most spectacularly with the detection of Earth-mass planets orbiting pulsars (Wolszczan & Frail, 1992). These early investigations culminated with the paper by Mayor and Queloz (1995) reporting a planet orbiting the solar type star 51 Pegasi, located about 50 light years away from the Sun. This latter planet was  found to be Jupiter-like, with a minimum mass of about 0.5 Jupiter masses, but with an orbital period of a 4.23 days equivalent to an orbital distance of 0.05 astronomical units. The discovery of a planet with these characteristics was unexpected by the scientific community, with the possible exception of Struve (1952) who first speculated on the existence of such hot Jupiters.

Since 51 Peg b was found with the radial velocity method (see below) which is biased towards finding massive planets in short period orbits, it was first written off as an unusual object. Soon thereafter Marcy and Butler (1996a, 1996b) discovered two more planets (7.4 and 2.4 Jupiter masses respectively) hosted by two solar analogues (70 Virginis and 47 Ursa Majoris) and with orbital periods of 116d and 2.98 years. None of these objects resembled planets found in our own Solar System and it seemed that giant planets with shorter orbital periods than in the Solar System were not so unusual. These first discoveries opened up the search for exoplanets that has now been going on actively for more than 20 years. Today, close to 2 000 objects are characterized as confirmed planets, while a further ~ 4 000 objects (detected by NASA's Kepler space mission – see below) are classified as Kepler Objects of Interest (KOI). The European CoRoT space mission, while confirming more than 35 planets, also has around 200 objects which are of interest but where the host star is too faint for proper follow-up and confirmation with current equipment. These have to await ongoing developments in instrumentation (e.g. with the ESPRESSO ground based spectrograph with first light due ~ 2016). As mentioned before, the first confirmed exoplanets were found orbiting around pulsars through radio pulse timing variations (Wolszczan & Frail, 1992). The formation process of these objects is not clear, but they presumably accreted out of debris produced by the supernova explosion that created the pulsar. Most exoplanets around normal stars have however, been discovered through one or more of three other methods, viz., the radial velocity method, the transit method or through gravitational lensing. All of these techniques were first applied from ground-based observatories and only later from space (gravitational lensing observations will be implemented on NASA's WFIRST mission and possibly also on ESA's EUCLID spacecraft). A detailed description of these methods can be found in Perryman (2011), or Haswell (2010). Briefly, the radial velocity method measures the radial velocity of the host star's reflex motion about the center of mass of the system due to the influence of the companion. Since it only measures the velocity component along the line of sight only the minimum planet mass is obtained ($m_p \times \sin i$ where $i$ is the orbital inclination relative to the line of sight to the host star). Assuming a value for the stellar mass, we get an estimate of the minimum mass of the planet, but not its true mass. The transit method is also very simple in that it searches for exoplanets whose orbital inclination is such that the planet passes between us, and the disk of its host star, causing the stellar flux to drop periodically. While the radial velocity method gives us the (minimum) mass, the transit method provides the radius of the exoplanet in terms of the stellar radius (which usually is unknown by factors > 10%). Because for transiting planets we are viewing the orbit nearly within the orbital plane, the orbital inclination is nearly 90 deg (and can be measured more precisely by fitting the transit light curve with an appropriate model). This means that radial velocity measurements now give us the true planetary mass, and consequently its average density can be calculated. As will become clear, the accurate measurement of ages, radii and masses of large numbers of exoplanets of different types represents the next advance in exoplanetology. The problem with the current data is that the accuracy of the planetary parameters of mass and radius are fully dependent on the accuracy of the same parameters for the host star. The current errors in the host star values are of order > 20%, >20%, and 50%-100% for the radii, masses and ages respectively. This will be further discussed below in the context of the TESS and PLATO future space missions.

**CoRoT and Kepler.**

The scientific case for a space mission to detect small transiting planets orbiting other stars began in earnest around the first half of the 1980s (see Roxburgh, 2011, Borucki, 2010 and references therein). This work led eventually to the launch in December 2006 of the European- Brazilian space mission CoRoT (Convection, Rotation and exoplanetary Transits; Baglin & Fridlund, 2006). CoRoT was the first space mission dedicated to the detection of exoplanets by searching for exoplanetary transits around solar-like stars. CoRoT utilized a relatively small (30cm) telescope with a limited field of view (4 deg$^2$). Due to its low-Earth orbit (900km), the longest the spacecraft could remain pointed towards the same target field was about 160d. Observing more than 20 individual fields, CoRoT, during a 6 year mission life time (nominally 2 ½ years), found more then 35 new planets (now published with another 10-15 being currently under confirmation) and acquired more than 160 000 light curves. CoRoT was the second spacraft designed for asteroseismology that was successfully launched and increased significantly the number of, and the precision with which asteroseismology is carried out from space (The Canadian microsatellite MOST was the pioneer in this context). CoRoT's planet yield include the first transiting terrestrial planet (Legér et al., 2009) that could be confirmed with follow-up radial velocity measurements of its host star (Queloz et al., 2009). This planet, CoRoT-7b has a diameter of 1.58 Earth-radii and a mass of 7.42 Earth-masses (Hatzes et al., 2011). The average density is of the same order as the terrestrial type planets in the Solar System. CoRoT-7b however, has a very unusual orbit. It orbits its G9V primary in 20.2h, clearly an object that has no analogue in our own system.

The next fundamental step was taken by NASA's Kepler mission launched in 2009. As a planet search mission Kepler currently holds the record with 1028 confirmed planets. Kepler consists of a one-metre aperture telescope that monitored about 160000 stars (about the same number as CoRoT) in a fixed 110 deg$^2$ field of view in the sky for a duration of approximately 4 years. The many thousands of transit candidates still require ground based (mainly radial velocity) follow up observations. Kepler was able to follow a significantly larger number of brighter stars than was possible for CoRoT. This allowed researchers to carry out asteroseismology on number of host stars and generally it was easier to make follow-up radial velocity observations of stars with smaller planet candidates. In May 2013, a second reaction wheel onboard Kepler failed, making it impossible to stabilize Kepler to the precision required for its initial mission. Since all other systems on board were working NASA issued a request for White Papers suggesting ways to use the space craft in its more limited mode. As a result of this, the K2 mission was developed.

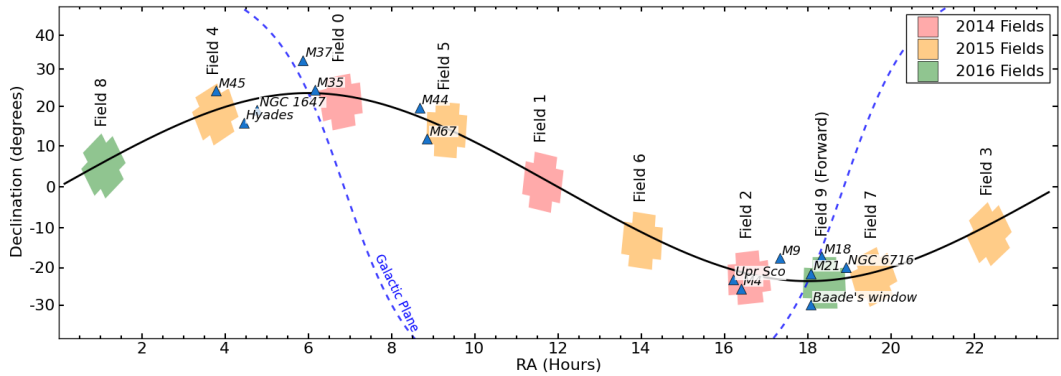

*Figure 1. The target fields for the 2-year K2 space mission.*

By observing in the ecliptic which minimized the force from the solar wind on the spacecraft and using its thrusters to compensate for the lost reaction wheel Kepler could continue to collect data. In this mode K2 observes targets in the plane of the Earth's orbit and in a direction away from the Sun. K2 can observe each target field for a maximum of 75 days before re-positioning for the next individual target field. This allows it to focus on bright stars and search for transiting planets that can be well defined in follow-up observations (see below). The K2 mission started in June 2014. Test observations demonstrated that a photometric precision of 50 ppm can be achieved for a 12 magnitude G-type star in 6.5h. All K2 targets are proposed by the community through the Guest Observer Program and the data is immediately available to all interested parties. Although K2's inheritance is based on exoplanetary research, the K2 mission welcomes all proposals including, but not exclusive to, exoplanet, stellar, extragalactic and solar system science. The mission is funded for 2 years, and will observe a total of about 10 fields (each 100 deg²) for 80 days. Each field will have about 10 000 targets. The selected fields are shown in Figure 1 and it is immediately noticeable that several important (in terms of stellar evolution) clusters will be observed.

Although the original Kepler mission is over, the analysis and follow-up of the material from its 4 year targeting towards the 110 deg² field in Cygnus is ongoing, as evidenced by the recent announcement of a super-Earth orbiting a G2V star within its habitable zone.

**The important next step.**

The conclusion that one can draw from these two first exploratory space missions is that the brighter the host star, the better one can constrain the exoplanetary parameters, and more precise work both as what concerns exoplanetary physics and the physics of the exo-atmospheres, can be carried out. And as has been shown (e.g. Rauer, 2014), high precision measurements of the planetary physical parameters are necessary in order to take exoplanetology to the next step.

The last 20 years have been very exciting when considering the discovery of literally thousands of exoplanetary systems. One of the most fundamental discoveries central to this wealth of data has been the realization of the enormous diversity displayed by these planets and systems. Whole new classes of planets have been discovered, orbiting in both very strange orbits and being of new types – compared to our own Solar System. An example demonstrating this is the aforementioned CoRoT-7b which is clearly a super-Earth with an Earth-like density. In the same system radial velocity observations have also demonstrated the presence of another Super-Earth in a 3d orbit, as well as a Neptune size planet orbiting in a 9d orbit and exterior to the other two planetary bodies. The discovery of similar objects (e.g. Kepler-10b; Batalha et al, 2011) thus indicate the existence of a class of objects much different to our own Solar System. The radius of CoRoT-7b was determined using the best possible photometric data (from space). Currently, the largest error on the density comes from the mass determination based on radial velocity measurements. The star CoRoT-7 may be bright by CoRoT standards, but it is relatively faint for radial velocity measurements. However, even if we could measure the radial velocity amplitude of CoRoT-7 with a very small error we would still not know if the structure of the planet is like (density wise) Mercury, or the Earth (Hatzes et al 2011) since we do not know the 'true' mass of the star CoRoT-7 itself. The same is true with the Kepler-78b object. This approximately Earth-sized planet orbits its host in a mere 8 hrs (Sanchis-Ojeda et al. 2013). Radial velocity measurements have established a planet mass of about one Earth mass (Pepe et al. 2013; Howard et al. 2013). Again, with zero error on the radial velocity amplitude we would still not know if the structure of this planet was like the Mercury, the Earth, or the Moon (Hatzes 2014) since we would not know the accurate stellar mass. The problem is that the ultimate planet radius and mass hinges on the accuracy of the stellar mass and radius. If we are to do precise exoplanetology to the point where we know whether a planet has a structure like the Moon, Earth, or Mercury we need to know the stellar parameters to a much better accuracy than we know now. The new instruments and satellites described below are in part intended to remedy this by making more exact observations of mainly transiting planets where both radii and masses can then be determined. Nevertheless, we will get both these parameters expressed in terms of the masses and radii of the host stars. And these are not known to a precision higher than maybe at best 10% which will introduce errors of the same order into the planetology. Our only way forward is to implement the new technique of asteroseismology, as intended with PLATO (see below) and as has been exquisitely demonstrated by Kepler (e.g. Metcalf et al., 2014). It has been shown (e.g. Rauer et al., 2014) that asteroseismology can provide an accuracy of 1-2% in radius, of better than 5% in mass and around 10% in age for solar-type main sequence stars as long as the photometric precision is good enough and the cadence is high.

# The Near Perspective: Ground

## Radial Velocities

At the moment, the "gold standard" in radial velocity measurements is held by the two HARPS (High Accuracy Radial velocity Planet Searcher) spectrographs, HARPS-N deployed at the Telescope Nazionale Galileo (TNG) in La Palma, Canary Islands, Spain. It is a 3.58m Alt-Az telescope equipped with an active optics system. HARPS-S has been in regular use since 2003 at the 3.6m telescope at the European Southern Observatory's La Silla Facility in Chile. These ultra-stable Echelle spectrographs, covering the spectral range of 380 to 690nm with a spectral resolution of up to R = 115 000, have claimed a radial velocity precision of around 0.5 m s$^{-1}$ under the best circumstances.

La Palma hosts the 2.6m Nordic Optical Telescope, equipped with the FIES (FIbre-fed Echelle Spectrograph) with a resolution of up to R = 67 000. By obtaining simultaneous Thorium-Argon spectra a very high stability is achieved and a radial velocity measurement precision better than 10 m s$^{-1}$ has been obtained.

Similarly, to the NOT/FIES combination, the ESO VLT/UVES combination can achieve relatively high spectral resolution (~ 5 m s$^{-1}$), and because of the aperture (8.4m) for significantly fainter stars. The same is true for the 10m Keck/HIRES combination, used with an iodine absorption cell as a wavelength calibration that achieves about 2-5 m s$^{-1}$.

The Automated Planet Finder Telescope (APF, commissioned in 2013) is a robotic 2.4-meter optical telescope at the Lick Observatory in California USA. It is designed to search for exoplanets in the 5 to 20 Earth mass range. It does so by performing high cadence observations on a sample of relatively few bright stars. A small number of bright stars are observed every night for 6 months. Although this telescope has a relatively small aperture, it is competitive in terms of the number of observations possible on a dedicated telescope. The philosophy is that it requires a large number of observations to beat down the stellar noise in order to detect really low mass planets.

All of these telescopes deployed today are either relatively small and therefore the magnitude range available for exoplanetary work is limited, or, when a large aperture is present, the spectrographs have not been designed with the ultra-stable requirements necessary in mind.  Further, the spectral range has so far been limited to the visual and therefore not very suitable for the searching and study of small (super Earth size) planets orbiting red dwarf stars that are both faint and very cool (e.g. Scalo et al., 2006). This latter requirement to study specifically red dwarf stars in this context has been introduced by the realization that if a red dwarf star would be a host to a Super-Earth planet orbiting within its Habitable Zone (HZ), it would be possible to characterize its atmosphere spectroscopically with already existing (or soon to be commissioned) equipment. There are, some interesting instruments delivering significant performance improvements that will be available soon.

The first such instrument is CARMENES ( Calar Alto high-Resolution search for M dwarfs with Exoplanets with Near-infrared and optical Echelle Spectrographs which will be deployed at the 3.5m telescope at Calar Alto in Spain. Built by a German-Spanish consortium it will provide a spectral resolution of R ~ 82 000 in two bands: One from 0.5μm to 1.0 μm, and a second one from 1.0μm to 1.7μm. It is specifically going to survey at least 300 red dwarf stars during 5 years, searching for super-Earths (about 5 Earth-masses). For this purpose, 600 *clear* nights have been reserved for the instrument in the 5 years starting from "first light", an event that is expected at the end of 2015. The instrument is specifically designed to be able to measure the radial velocity of the target star in both wavelength bands. False positives due to stellar activity will manifest themselves with different velocity amplitudes in the optical and infrared.

The second instrument is ESPRESSO (Echelle SPectrograph for Rocky Exoplanet and Stable Spectroscopic Observations) that will be deployed at one of the 8.2m Very Large Telescopes (VLT) at the ESO Paranal facility in Chile. This ultra-stable spectrograph which will be sensitive between 350 and 720nm will have a spectral resolution of 140 000. The stated goal is to reach a precision of 0.1 m s$^{-1}$. First light is expected in 2016.

In order to further improve on the stability of spectrographs and thus the precision (eventually to better than 10 cm s$^{-1}$) the technology of Laser frequency Combs is being developed for radial velocity work (Lo Curto et al. 2012), but may not reach the maturity level for a usable instrument (in the visual wavelength range) before 2020. In the infrared, however, the recently funded upgrade of NIRSPEC on the Keck telescopes in Hawaii intend to implement laser combing as a reference, aiming for a velocity resolution of 1 m s$^{-1}$ and first light in 2016/2017 (Beichman, private communication).

**Transits**

On the ground a large number of new networks of telescopes searching for exoplanetary transits of smaller and smaller planets are being deployed. These include:

The "Next Generation Transit Survey" (NGTS) which builds on the previously so successful SuperWasp program that utilized robotic telescopes at two sites. WASP has a proven record of detecting transiting Jupiter size planets, whereas NGTS has as its objective the detection of down to Neptune size exoplanets. NGTS employs an array of robotic small telescopes operating in the 600-900nm band, thereby maximizing sensitivity to bright but relatively small and cool host stars (K and early-M spectral type). NGTS will survey the brighter stars that CoRoT and Kepler could not access. This will provide scores of prime targets for high quality characterization by the next generation of telescope/instrument combinations such as the VLT/ESPRESSO, the E-ELT and the space missions JWST and CHEOPS. Bright host stars provide the key element to this follow-up, particularly to the radial velocity measurements. But since the discovered exoplanets will all be transiting, the brightness of the host star will also enhance the possibility of investigating their atmospheres spectroscopically.

NGTS is located at a site with excellent photometric conditions, namely the ESO E-ELT site at Cerro Armazones, about 20 km away from the Paranal facility (location of VLT). The science requirement (0.1% photometric precision across a wide field) is more technically demanding than in the case of previous ground based transit surveys, and the key technologies has been demonstrated with a prototype system. The NGTS telescopes are 200mm, f/2.8 high-quality commercial telescopes and 12 such telescopes are equatorially mounted on individual piers. Each telescope has a field of view of 8 deg$^2$. The telescopes are working together, and the total field covered thus is 96 deg$^2$. Each telescope is equipped with red-sensitive (2k×2k) CCDs designed to observe with optimal sensitivity in the 600 to 900nm range in order to match the peak emission of the primary K and early M stars targets.

MASCARA, The Multi-site All-Sky CAmeRA is a unique concept to monitor the whole sky looking for exoplanets around the brightest stars in the sky (< 8 mag in V). Previous surveys like the aforementioned SuperWasp, as well as HATNET and TrES targeted stars with a magnitude fainter than V = 8 due to saturation limits on their detectors and exoplanets around the brighter stars have up to now only been detectable via radial velocity measurements. MASCARA is a fully funded network consisting of 5 stations across the globe, with each station composed of a battery of cameras to monitor the near-entire sky at each location. Once all stations have been installed, MASCARA will be able to provide a nearly 24-hr coverage of the complete sky, down to magnitude 8, at a sub-minute cadence. Its main purpose is to find the brightest transiting exoplanets, expected in the V=4-8 magnitude range, but by surveying so many bright stars MASCARA will also allow for a wealth of secondary science. The design of each station with its cameras is especially interesting. The basic principle is to be able to use modified CCD based cameras developed for the amateur astronomy market, mated to commercial camera lens systems. The concept has been tested and the required precisions have been confirmed.

**The Near Perspective: Space**

Kepler and CoRoT demonstrated that space is the best place for transit studies. In this respect the situation looks bright with two upcoming missions. After CoRoT and Kepler/K2, the next two space missions are TESS and CHEOPS, both of which will be studying the transit light curve of exoplanets. Both are going to follow the recommendation based on the results of their predecessors that in order to be able to study the physics of exoplanets we need to observe the light curve of transiting exoplanets orbiting *bright stars* in order to be able to carry out also the follow-up observations with the necessary precision.

**TESS:** The Transiting Exoplanet Survey Satellite (TESS), planned for a launch in 2017, is a NASA all sky survey to be conducted over 2 years. It will monitor the brightness of a total of ½ million stars and study 100 000 of the brightest stars in detail. It will map the northern half of the sky in the first year and switch to the southern half in year 2. The spacecraft will be using an array of 4 wide-field 10 cm cameras scanning the sky. Each camera features a low-noise, low-power 16.8 megapixel CCD detector with a 576 deg² field. Due to the scan pattern, most targets will be followed for about 30d. A small percentage will be followed for a longer period up to about 160d. TESS will focus on relatively bright G- and K-type stars where the transits will be excellent targets for further study with the James Webb Space Telescope (JWST). There will also be a prioritization on all available red dwarf stars. These stars have a habitable zone quite close to the star and radial velocity confirmation of rocky planets in this zone is supposedly relatively easy. TESS will also continue the observations of the photometric variations due to stellar oscillations ("asteroseismology") begun by CoRoT and Kepler. The result of those two pioneering missions indicate that TESS will detect p-mode oscillations (variations where the stochastically induced acoustical waves are restored by pressure) for about 6 000 of the brightest stars (brighter than $V = 7.5$). These objects include about 2 000 stars on the upper main sequence as well as subgiants, and all giant stars within the magnitude range. While TESS will offer opportunities to detect stellar variability for its (in total) 500,000 targets, its limited time base of 30 days will prevent the high-precision seismic age determination for stars with the exception of those located at the ecliptic poles.

**CHEOPS**, CHaracterising ExOPlanet Satellite is the first mission dedicated to searching for exoplanetary transits by performing ultra-high precision photometry on bright stars already known to host planets. The mission's main science goals are to measure the average density of super-Earths and Neptunes orbiting bright stars and thus to provide suitable targets for future in-depth characterisation studies of exoplanetary atmospheres. It is the first of ESA's "small missions", carried out together with a European consortium led by Switzerland, and is planned for a launch in 2018. The main difference between CHEOPS and previous exoplanet missions is that one will already know if a planet is present from radial velocity observations and one will likewise have a reliable ephemeris for when to expect the transit. The data will allow the determination of the densities of the planets studied, the errors only being dominated by the assumed/modeled host star mass and radius. CHEOPS will have the capability to measure a photometric signal with a precision of 150 ppm/min for a V=9 magnitude star. This allows the detection of the transit of an Earth-sized planet orbiting with a period of 60 days around a star of 0.9 $R_{sun}$ and with a S/N of the transit greater than 10. CHEOPS uses a single frame-transfer back-side illuminated CCD mated to a 32 cm on-axis Ritchey-Chretien telescope with the image of the target star being de-focused onto the detector. Special care has been taken to achieve a design that minimizes stray light entering the telescope and using a dedicated field stop and a baffling system. Calculations and testing demonstrate that the design meets the requirement of less than 10 ppm stray light falling onto the detector even in the worst observational scenario. Thermal stabilization of the detector to within 10 mK is obtained by radiating surplus heat to cold space. The mission critical technology of CHEOPS have already been space qualified since it was flown on-board the CoRoT space craft where the baffling/field stop combination surpassed the strict requirements by an order-of-magnitude in a similar orbit. On CoRoT, the 1-sigma noise levels reached less than 1 ppm per mHz$^{1/2}$ for the brightest stars (around 6:th magnitude). For a 11.5 magnitude star the noise level was measured to be 2.00 x 10$^{-4}$ per 512s integration (Auvergne et al., 2009). These numbers are probably indicative for the levels reachable by CHEOPS.

A summary of the main science objective of the CHEOPS mission is:

- The study of exoplanets with sizes ranging between one and six Earth radii and the following goals:
    - Deriving the mass-radius relation in a planetary mass range for which only a handful of data exist and with a precision not previously achieved
    - Identifying planets with significant atmospheres over a wide range of planetary masses
    - Placing constraints on possible planet migration paths followed during the formation and evolution of planets
    - Probing the atmospheres of known hot Jupiters
    - Providing unique targets for future ground-based (e.g. E-ELT) and space-based (e.g. JWST) observatories with spectroscopic capabilities.

**Astrometry and Gaia:**

Apart from when we observe a planet transit its host star, the orbital inclination and thus the true planet mass as discerned from radial velocity measurements is not known. And in the case of transiting planets the inclination is usually only known for small star-planet distances due to the lower probability of a transit occurring for planets with large orbital radii. This became an interesting complication when the first exoplanets were detected with the radial velocity technique and one thus had only a minimum mass of the planet itself. This led to interpretations where the planets could easily be considered to be over the limit (then usually stated as being 13 $M_{Jupiter}$) that would be transforming them into brown dwarfs (e.g. Latham et al., 1989).

A clear example of this problem is the object found orbiting the star HD 33636, where radial velocities gave a minimum mass M x sin i = 10.2 $M_{Jupiter}$. Astrometric measurements using the Fine Guidance Sensor (FGS) of the Hubble Space Telescope (HST) determined an orbital inclination of only 4 degrees which results in a true companion mass of 142 $M_{Jupiter}$ or 0.14 $M_{Sun}$ (Bean et al. 2007). Astrometric measurements with FGS have over the years been able to measure the true mass of a few exoplanets. These include GL 876b (Benedict et al. 2002), ε Eri (Benedict et al. 2006), and the brown dwarf companion around HD 136118 (Martioli et al. 2010). In the case of 55 Cnc astrometric orbits gave the mass and orbital inclination for two of the planets in the system, 55 Cnc c and d. The orbits of these two planets are mis-aligned having a mutual inclination of 29 degrees (McArthur et al. 2010). However, due to the limited sensitivity of performing such astrometric measurements with the FGS on HST such important observations have only been done in a very few cases.

This situation is about to change. In December 2013, after more than a decade of development, ESA launched its successor to the HIPPARCOS satellite (which flew 1989 – 1993), Gaia. The name was originally an acronym for Global Astrometric Interferometer for Astrophysics, reflecting its inheritance from ESA's program for Space Interferometry. Astrometry (GAIA) together with exoplanetology (Darwin) originated in ESA's Horizon 2000 program initiated by the then Director of Science, Roger Bonnet, that identified Interferometry carried out from Space as a so-called 'Green Dream' topic, i.e. something clearly to be carried out when technology had matured enough. However, the first studies of GAIA carried out industrially demonstrated clearly that it would be much more (cost-)effective to base the new spacecraft on the principles used already for the previously successfully flown HIPPARCOS, taking advantage of advances in technology but without the added complexities of interferometry. It was demonstrated that goal to achieve 2 – 3 orders of magnitude improvement in astrometric precision as well as observing more than 1000 times the number of stars than what was achieved by HIPPARCOS, was feasible. Thus the name of the mission changed to Gaia. Gaia has thus begun to take a census of roughly 1 billion stars or about 1% of all stars in the Milky Way Galaxy. The spacecraft has a nominal lifetime of 5 years and during this period each star should be observed about 70 times. During each observation the position, brightness and color are recorded. From these data, the stars distance and proper motions can be calculated with an unprecedented precision.

Gaia has three instruments. The most important is the astrometric that records the passing of a star across a grid projected onto the hitherto largest CCD assembly flown in space instrument that as the spacecraft rotates. The second instrument is a spectrograph that records the radial velocity of each star. This third dimension together with the x- and y- position on the sky will provide a three-dimensional map of the Galaxy. A third instrument will measure the color of the stars observed so as to derive effective temperatures. Gaia will study exoplanets in several fashions. It will discover giant exoplanets by detecting the wobble in the plane of the sky that the orbital motion of the exoplanet induces in its host star. Although it makes relatively few photometric observations, by virtue of the large number of stars that will be observed Gaia will also be able detect some transits of planets as they pass between us and their host star (see Perryman et al. 2014). Finally Gaia may also be able to confirm massive giant planets and brown dwarfs through the radial velocity measurements. Gaia will also add to the exoplanetary science by providing an "input catalogue" for ESA's M3 PLATO mission (see above). Gaia's highly precise stellar distances of all of future PLATO target stars will also greatly help that later mission by improving the asteroseismic data that PLATO will obtain through the removal of some of the properties of the exoplanetary host stars as free parameters. Accurate distances, magnitudes, and effective temperatures can fix the stellar radius of the object in the retrieval of the asteroseismic parameters. This will provide an increase in the precision planet's physical data.

**The Next Steps: Ground**

**E-ELT (2024):**
At the moment several Extremely Large Telescopes are being considered or are under development worldwide as one of the highest priorities in ground-based astronomy. While such telescopes will also study and vastly advance astrophysical knowledge in many disciplines such as the first objects in the Universe, super-massive black holes, and maybe contribute to the issue about the nature and distribution of the dark matter and dark energy what we are most concerned with here is the major advances that will be possible concerning planets around other stars. Of these instruments the E-ELT (European Extremely Large Telescope), being developed by the European Southern Observatory (ESO) this revolutionary new ground-based telescope concept will have a 39-metre main mirror and will be the largest optical/near-infrared telescope in the world. The E-ELT program was approved in 2012 and green light for construction was given at the end of 2014. First light is targeted for 2024. One of the stated goals for this instrument is "to track down Earth-like planets around other stars in the "habitable zones" where life could exist". While the detailed calculations showing what can be done in this context is lacking, it is clear that the aperture alone of this instrument will allow the detection of enough photons. Nevertheless there are significant problems to be clarified. While it is clear also that the telescope can in principle resolve the 0.1 arcsec required to separate the Earth from a G2V star at 10 pc, the issue of how to correct for speckle noise in a telescope with many hundreds of segments at the relevant timescales in order to diminish the contrast has to be clarified. Spectroscopy appear more directly feasible at this time, either as what concerns radial velocity detection or the direct detection of spectroscopic features in exoplanetary atmospheres. It of course assumes that a suitable spectrograph is made available at the appropriate time.

**The Next Steps: Space**

**JWST (2018):**

The James Webb Space Telescope (JWST) observatory consists of a passively cooled telescope that is optimised for diffraction-limited performance in the near-infrared ($2\mu m$–$5\mu m$) region, but with extensions to either side into the visible ($0.6$–$2\ \mu m$) and mid- infrared ($5$–$28\mu m$). This is in contrast compared to the Hubble Space Telescope's (HST) $0.1\mu m$–$2.5\mu m$ (ultraviolet to the near infrared) wavelength range. The JWST observatory includes three main elements, the Integrated Science Instrument Module (ISIM), the Optical Telescope Element (OTE) and the Spacecraft Element which comprises the spacecraft bus and the sunshield. The primary mirror is 6.5 m in diameter and consist of 18 mirror elements made of gold-coated beryllium It will have a giant sun-shield (22 x 12 m) protecting the telescope and the instruments from the light and heat of the Sun. The total mass of the JWST will be 6,500 kg and the spacecraft will be launched with a European Ariane 5 rocket into a Lagrangian L2 orbit located approximately 1.5 million km away from the Earth. This makes its operation and pointing/stability requirements both much simpler and more efficient, in comparison with Hubble. Of course refurbishment missions to JWST will be impossible in this orbit.

The science objectives of the JWST falls into several broad categories, *viz.* the observation of:

- The phase of the evolution of the Universe when the first generation of stars were formed and started to shine, thereby re-ionizing the Universe.
- The first formation of early galaxies and clusters of galaxies
- Detailed observations of star formation and the accompanying process of planetary system formation
- "Planetary systems and the origin of life" taken to mean the first high sensitivity spectroscopic observations of exoplanetary atmospheres, as well as complex molecules in the interstellar medium accompanying the star formation process

### **Millimetron (2019/2022?):**

The 12m MILLIMETRON is a radio astronomical observatory that will be launched by Russia and is equipped with Bolometers operating in the 200μm – 400μm and 700μm – 1400μm ranges, having a sensitivity at 300μm and 1 hour integration that is 1 nJy. It will also be equipped with 3 spectrometers for the 50μm – 300μm ($R = 10^6$), 30μm – 800μm ($R = 1\,000$) and 20μm – 2000μm ($R = 3$). The idea is also to use it as an interferometer (like its predecessor Radioastron), using other radio telescopes on the ground as the other nodes and creating a very long baselines.

### **PLATO (2024):**

After CHEOPS and TESS, Europe will return to the study of exoplanets from space with the PLATO mission which will take the results of those missions to the next level. The PLATO spacecraft (Rauer et al. 2014) was selected, in February 2014, as the European Space Agency, ESA's, medium class mission for the third (M3) launch opportunity upcoming in 2024 in the Cosmic Vision program.

It addresses very fundamental questions relating to the study of exoplanets such as:

1. How do planets and planetary systems form and evolve?
2. Are there other systems with planets like ours?
3. What kind of planets do we find within the Habitable Zone around different types of stars?
4. What is the role of the host star in the context of the formation and evolution of exoplanets?

PLATO is based on the experiences of both the CoRoT and Kepler missions, and consists of a spacecraft bus, equipped with 34 individual 12 cm aperture telescopes, thus providing both a wide field-of-view, yet a large photometric magnitude range.

Two of the telescopes are equipped with red- and blue-filters and are used both for the pointing and the stability control, as well as in order to observe the very brightest of the target stars, giving the system as a whole a very large dynamical range. PLATO targets bright stars, primarily in order to detect and characterize planets down to Earth-size or smaller by photometric transits and whose masses can then be well determined by ground-based radial-velocity follow-up measurements. PLATO will be the first mission with sufficient time base, precision, and cadence to derive masses, radii, ages, and internal rotation profiles with very high accuracy for 100,000s of stars across the Hertzsprung-Russel Diagram utilizing the new tool of asteroseismology (e.g., Aerts et al. 2010 with updated reviews in Chaplin & Miglio 2013 and Aerts 2015). The asteroseismology will be used for the determination of the stellar fundamental parameters for all stars brighter than about magnitude 11.5. The combination of bright targets and asteroseismology results in high accuracy for the planet parameters, quantifiable as better than 2 %, 4–10 % and 10 % for planet radii, masses and ages, respectively. The foreseen baseline observing strategy includes two long pointings (2–3 years each) to detect and characterize planets down to and beyond Earth-size and reaching into the habitable zone (HZ) of solar-like stars, pushing the border beyond previous missions. An additional so-called step-and-stare phase with shorter pointings (1 – 5 months each and in special cases just a few days to confirm earlier long term detections) is filling up the remaining lifetime of the spacecraft. It is estimated that the survey will cover at least about 50 % of the sky during the nominal 6 year mission. For a potential extended mission of up to 8 years and beyond it is possible to cover more than 70–80 % of the sky. During the nominal mission PLATO will observe up to 1,000,000 stars and will discover and characterize hundreds of planets in the Earth to Super-Earth category, and thousands of planets in the Neptune to gas giant regime in orbits out to the HZ, as well as detecting many more for which partial characterization will be possible. PLATO will therefore provide the first large-scale catalogue of planets with radii, masses, mean densities and ages, accurate enough to carry out a detailed analyses of each planet. This catalogue will include Earth-like planets at intermediate orbital distances, where surface temperatures are moderate. Coverage of this parameter range, with a statistical large number of characterized planets, is unique to PLATO. PLATO will provide a census for small (low-mass) planets. The results of the mission will help complete the knowledge of planet diversity for both low- and high- mass objects (at orbital distances < a few AU), and correlate the planet mean density as a function of orbital distance. These parameters can be compared to predictions from planet formation theories. Other investigations that are possible to carry out include the constraining of theories for planet migration and scattering, and to specify how planet and system parameters change with host star characteristics, such as stellar mass, metallicity and age. The catalogue will also allow us to study planets and planetary systems at different evolutionary phases. This will serve to identify objects which are retaining their primordial hydrogen atmosphere and in general the typical characteristics of planets in the low-mass, low-density range. Since PLATO observes mainly bright stars, many of the discovered planets will be targets for future spectroscopic investigations exploring their atmospheres. Furthermore, the mission has the potential to detect exo-moons, planetary rings, binary planets, and Trojan planets.

The planetary science possible with PLATO is complemented by its impact on stellar and galactic science. Asteroseismology will provide us with fundamental stellar parameters including stellar ages for a large number of stars spread throughout the Galaxy. Precision light curves will be obtained for all kinds of variable stars, including those in stellar clusters of different ages. This will allow us to improve stellar models and to have a better understanding of stellar activity as a function of stellar age. A large number of well-known ages and masses of red giant stars will probe the structure and evolution of our Galaxy. Asteroseismic ages of bright stars for different phases of stellar evolution will allow us to calibrate stellar age-rotation relationships. Together with the results of ESA's Gaia mission, the results of PLATO will provide a huge legacy to planetary, stellar and galactic science. Most importantly, with its high accuracy and sensitivity, PLATO will make the first unique observations that will take the field of exoplanetology from its infancy to a mature phase where we will be able to carry out geophysics on Earth-like worlds and compare these observations with data from our own Solar System. At the same time, observations of the host stars will give us accurate ages for the exo-systems, allowing us for the first time to study the evolution of planetary systems and thus understand the context of our own world among these systems.

**The further future**

**SPICA (Beyond 2025?):**

SPICA, The Space Infrared Telescope for Cosmology and Astrophysics is a 3.5 m cryogenically cooled space telescope proposed by the Japanese community through its space agency JAXA/ISAS and intended to be carried out in collaboration with ESA and NASA. It has been considered to be equipped with a coronograph for high resolution, high dynamical range observations of disks and exoplanets, as well as being equipped with high- and low-resolution spectrometers in the 4µm to 210µm wavelength range. This will complement the wavelength ranges studied with JWST as well as Herschel (which are/were not cryogenically cooled) bringing high spatial resolution to the cool Universe (20K-40K).

The telescope has also been considered to be cooled to below 6 K allowing unprecedented sensitivity in the long wavelength infrared regime. SPICA would then be an essential pre-cursor to more futuristic facilities such as the proposed large space based infrared interferometers as it would demonstrate many of the technologies required for these missions. SPICA should have a guaranteed three year lifetime, but, because the cooling of the telescope and instruments is achieved using closed cycle mechanical coolers, it would be expected that the actual operational life could be significantly longer than this. SPICA's status at the moment is that it is awaiting a new proposal to the relevant space agencies.

The instrument suite for the SPICA satellite could comprise the following instrument capabilities:

**Mid-infrared camera and spectrometers:** 5µm-37µm imaging and spectroscopy. The grating spectrometers will have a resolving power of typically a few thousand and will cover the full MIR including the 27μm - 37μm band not accessible to JWST. As an option there would also be two immersion grating spectrometers operating in the short wavelength MIR (5-15 μm) with resolving powers of up to 30000. These would allow, for instance, the study of the dynamics of protostellar disks.

**Far Infrared Spectrometer and Camera: Utilized for** 34μm-210μm imaging and spectroscopy. This instrument will be a Mach-Zehnder configuration Fourier transform spectrometer to provide imaging spectroscopy from 34μm-210μm covering from the rest frame [Si]34μm to [NII]205μm lines over a 2x2 arcmin field of view.

**Guidance Camera:** NIR and visible focal plane camera with a scientific imaging and low resolution spectroscopy capability.

**MIR Coronagraph:** 5-27 μm coronagraph with medium to low resolution spectroscopy. This instrument will have an inner working angle of 3.3λ/D allowing it to image and take modest resolution (R = 10-200) spectra of young gas giant planets orbiting as close as about 10 AU around stars at 10 pc distance. There is also an option to extend the wavelength range down to 3.5μm.

SPICA would have the capability to directly detect and undertake detailed characterisation of a number of hot young gas giant planets using its coronagraph. Additionally the MIR spectrometers will be capable of undertaking transit spectroscopy on many targets with high efficiency and high sensitivity. The coronagraphic capabilities of SPICA could, in principle, exceed those of JWST as the PSF will be very much cleaner and the dedicated instrument would have a rather better contrast ($10^{-6}$ compared to $10^{-4}$ for JWST). A unique feature of the coronagraph would be the ability to take direct spectra of a few target exo-planets. The far infrared region is unexplored territory for exo- planet research and it is anticipated that both transit photometry and, possibly, spectroscopy will be possible on some targets allowing us to probe to different depths into planetary atmospheres and opening a new discovery space in characterisation.

Also directly related to exo-planets, the detection and characterisation of exo-Zodi dust clouds is one of the prime goals for the SPICA FIR instrument. It will follow up on the results of Spitzer and Herschel, and will allow a much more complete picture of planetary formation scenarios and the role of planetesimals.

## FIRI (2030?):

The European Space Agency, ESA, first began to look into a Infrared technology interferometer in the context of the nulling interferometer for the Darwin study in 1997. In 2006, a study began of a more general purpose instrument based on constructive interference in the Infrared wavelength region. The Far InfraRed Interferometer (FIRI) Technology Reference Study investigated the feasibility of an interferometer to achieve an angular resolution of less than 1 arcsecond at wavelengths between 25µm and 300µm. The selected baseline concept was based on a single spacecraft with two cryogenically cooled telescopes on two deployable booms and with a central hub, forming a Michelson interferometer. This would be quite different to the previously studied 'nulling' interferometer Darwin, which consisted of 3 – 4 telescopes (passively cooled), and free-flying in space allowing very large baselines. FIRI would address the same wavelength region as SPICA, but with significantly higher spatial resolution.
A later study (http://arxiv.org/pdf/0707.1822.pdf), utilized the Darwin concept with instead 3-4 telescopes and with the telescopes and the beam-combiner in free-flying mode.

## Exoplanet and disk studies in the future

To continue the work on star- and planet-forming disks, on debris disks, and the equivalent of the zodiacal- and Kuiper-Edgeworth disks it appears that a number of possible important instruments are either already present (ALMA), in development (JWST and the E-ELT) or suggested (SPICA, or FIRI). The importance of high signal-to-noise, high spatial resolution spectroscopic IR data can not be underestimated in this context, but the situation appear to be well in hand.

However, one of the more important problems to be addressed in the long term time schedule concerns direct spectroscopy of habitable planets with the capability to detect signs of biological activity at interstellar distances, and here the situation is more unclear. One may argue that it is to early to start designing such an instrument, since we do not know, at the moment, exact what (type of) target(s) to design for. Either Super-Earths orbiting in the HZ around a very late M-dwarf, or the a 'normal' Earth analogue orbiting at 1 AU around a G2V star would require tow completely different instruments, particularly considering that the former would likely be at a distance less than 5pc (in order to be discovered at all), while the latter would very likely be much further away (there are only a handful of G-type main sequence stars within 10pc; Kaltenegger et al, 2010). Regardless of which type of target will be the first objective, it is a formidable problem, since, that while a reasonable spectral resolving power for measuring the spectrum of an Earth-sized planet in a habitable orbit with a V-mag of 35 is only 10 -100, the main problem is one of contrast and spatial resolution. The host star is >$10^{10}$ brighter than the planet in the visual, and located with a spatial separation of < 0.1 arcsec (0.1 arcsec at 10pc for a planet orbiting a G2V star at 1 AU separation and progressively much closer for more distant objects or for fainter host stars). This can be achieved either through imaging – with or without coronography – or through (destructive, so-called 'nulling') interferometry. A third option being pursued mainly in the USA is the "Starshade" or "Occulter", which is basically a variety of coronography but where the mask is flown in space at a great distance (many tens of thousands of km) away from the actual telescope.

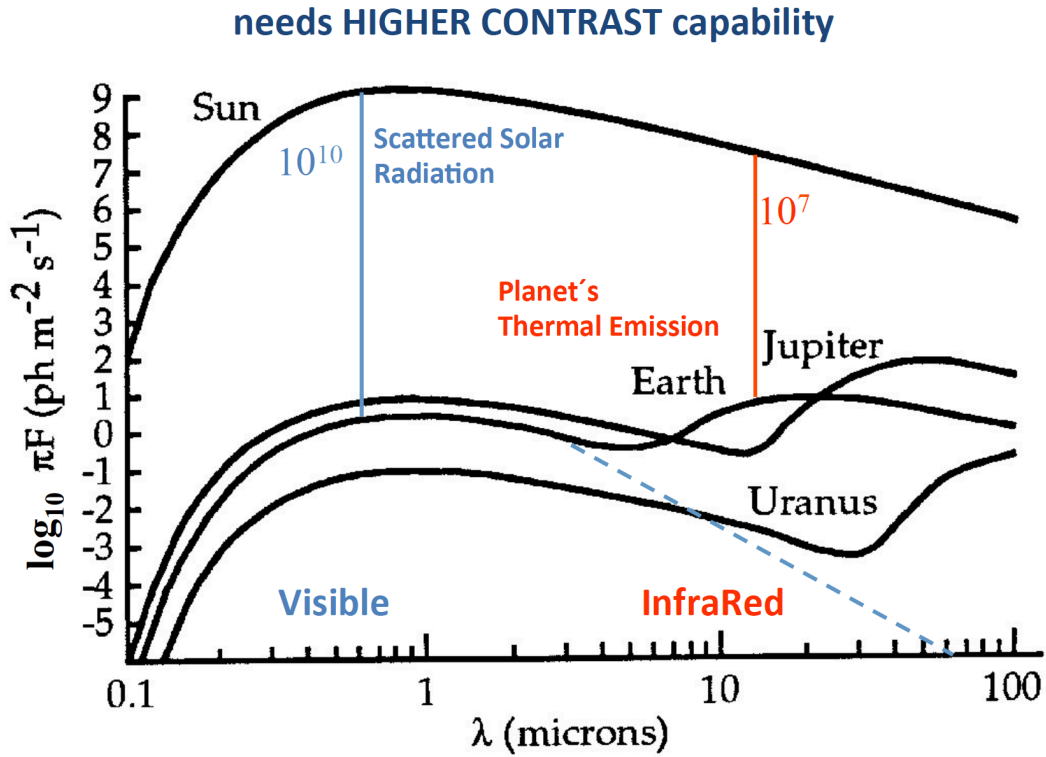

Figure 2: The contrast at different wavelengths for the case of the Sun and a few of the Solar System planets

In obtaining a spectrum of a habitable exo-earth one needs to choose to observe the appropriate region of the electromagnetic spectrum. The contrast is several orders of magnitude lower if one selects the near- to mid-infrared spectral region and scattered light from the host star will be (somewhat) less of a problem (see Figure 2). At the same time, in order to achieve the required spatial resolution, one requires larger optical elements in the infrared, which have historically been one of the drivers for using an interferometer in space. Finally and ultimately, obtaining spectroscopy of a "habitable" planet should indeed also demonstrate not only if it is indeed 'habitable' but also whether it is inhabited by life forms. One need thus to determine and select so-called 'biomarkers' in the spectrum that can be observed in the appropriate wavelength range. With this criterion one has entered the field of Astrobiology (Figure 3).

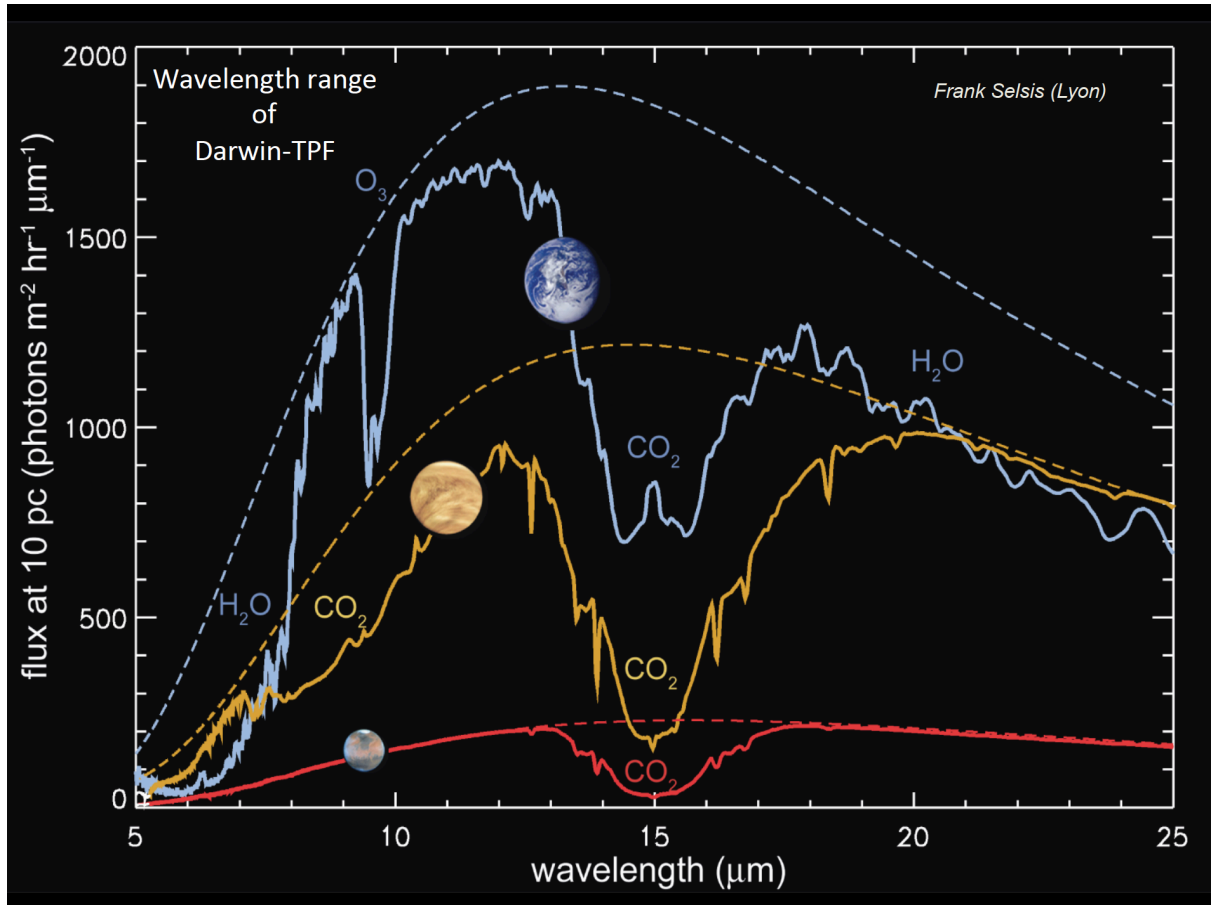

*Figure 3: An example of spectral features that could be considered to be 'biomarkers' – at least in our own Solar System*

Taking all these complications into account, it appears that the development of the different technologies required should be started now. The development time of flight hardware is so long that it will be barely ready in the time when we are likely to have identified suitable targets. And the space agencies of the world have realized this long time ago. In 1996 ESA and NASA began parallel but coordinated studies of how such an instrument could be defined, developed and deployed in the shortest possible time. The ESA study went under the name of Darwin, while the NASA effort was designated Terrestrial Planet Finder or TPF for short.

Although suggestions for a possible technology, as well as a definition of a realistic scientific case had been put forward much earlier (e.g. Angel, Cheng & Woolf, 1986; Burke, 1988; Bracewell & MacPhie, 1979), this was the first time the problem was studied systematically by the worlds two largest space agencies and in conjunction with the worlds leading aerospace industries. At the time – 1996 – only 2-3 exoplanets had been found and were still questionable in many scientists mind. Particularly since the newly discovered planets were of types (e.g hot Jupiters and sometimes with wildly eccentric orbits) not present in our Solar System. The Darwin and TPF studies were therefore targeted towards clones of our own Solar System, i.e. early G-type stars orbited by terrestrial planets in the Habitable zone, with gas giants in the region beyond (later, the Kepler mission was targeted towards the same objective). The design specifications were thus aimed towards detecting a 1 Earth-radii planet orbiting 1 astronomical unit away from a G2V star. Between 1996 and 2007 ESA and NASA developed space components that could detect such a system out to 33pc distance. The ESA study was based on a so-called nulling interferometer, while NASA also studied the Coronography and eventually also the occulter options. In 2000 ESA decided to continue the technology development of the nulling interferometer and this work progressed until 2007 when it was closed down. By then ESA had validated all technologies required in a free flying interferometric telescope based on destructive interference (Figure 4). Specifically the free-flying of a constellation of spacecraft, based on the "Darwin-technology" was also later space qualified by Sweden in the PRISMA experiment that was free flying two nano-satellites, utilizing this technology, in low earth orbit.

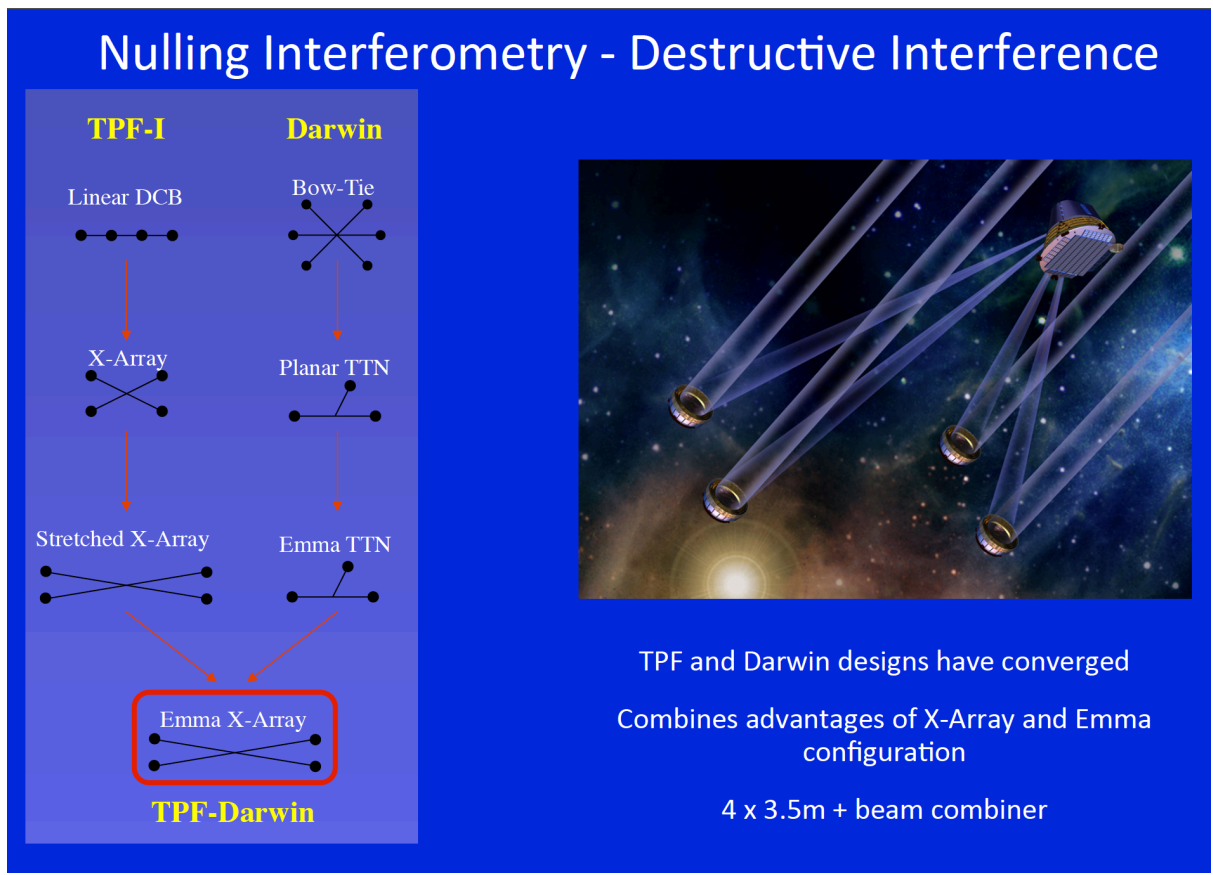

*Figure 4: Different constellation concepts evaluated in ESA's Darwin study, as well as an artists rendering of a 4 telescope array plus beam combiner. The telescopes in this array are passively cooled 3.5m telescopes.*

Following the definition and technical studies of these major mission concepts during the period 1996–2007, very little activity has taken place in Europe. The final conclusion of the studies were that several technologies exist with the capability of studying the direct light from small planets in habitable orbits around stars at least out to a distance of 30–50 pc. Which technology to use, depend, however, on the target in question. After 2007, some work has been carried out, mainly in the US, but the extensive technological effort aimed at a launch anytime soon has slowed down significantly. There are some studies being carried out w.r.t coronography by NASA. This is also true for the so-called New Worlds Missions, where studies of the deployment of star shades, i.e. external occulters intended to fly many tens of thousands of km away from the telescope have been carried out utilizing literally hundreds of nano-satellites brought together in orbit to assemble the large structures required (Seager, *private communication*).

At the moment exoplanetology has reached an impassé. In spite of the more than 4 000 exoplanet candidates we currently have in our catalogues, none is exactly like our own Solar System. A handful of super-Earths have been tentatively identified in what could be designated as 'habitable' orbits around the stars in question and even one case, Kepler 452b (Jenkins et al., 2015) with such a planet at roughly one AU from a 'true' Solar analogue. Since essentially all objects for which we have data both on the radie and the mass tend to be located many hundreds of parsecs away from the Earth (the CoRoT and Kepler planets) they would not be observable with the Darwin or TPF interferometers/coronographs (typically too small separation between host star and planet). Therefore, we at this moment do not know what kind of instrument to select for the objective of obtaining direct spectroscopy of Earth analogues. Even if such planets turn out to be relatively 'common' the observable targets may be few since most may be located at large distances. The instruments – especially if it is necessary to deploy them in space – will likely turn out to be complicated, long-term and expensive. It is therefore absolutely required that we continue the present work that has been described in this chapter in order to eventually identify an as large as possible assemblage of targets before we commit to the appropriate instrument. Regardless of if this then turns out to be an interferometer, a coronographic large space telescope or a system with a free-flying occulter, the scientific case is strong enough to motivate a great effort.

Luckily, the situation with the new swath of instruments, new networks of telescopes, the E-ELT and its American cousins, CHEOPS, TESS and PLATO being either under development or selected give a clear promise that such Earth-analogue targets will be identified within the near future. The design and implementation of the appropriate instrumentation for the detailed characterization of their properties, especially the habitability, may then go ahead.